\documentclass[twocolumn,SC]{revtex4}
\usepackage{graphicx,float}
\usepackage{amssymb,amsmath}

\begin{document}
\title{\bf Ginzburg-Landau Formalism in a Tilted Dirac Cone Metric}
\author{ M. A. Rastkhadiv \footnote{E-mail: rastkhadiv@shirazu.ac.ir}}
 \affiliation{Estahban Higher Education Center-Shiraz University, Estahban, Iran}

\begin{abstract}
Recent researches on tilted Dirac cone materials have unveiled an astonishing property, the metric of the spacetime can be altered in these materials by applying a perpendicular electric field. This phenomenon is observed near the Fermi velocity, which is significantly lower than the speed of light.  According to this property, we derive the Ginzburg-Landau action from the microscopic Hamiltonian of the BCS theory for the tilted Dirac cone materials. This derivation is performed near the critical point within the framework of curved spacetime. The novelty of the present work lies in deriving a general Ginzburg-Landau action that  depends on spacetime curvature, where the curvature is tuned by an external electric field. Furthermore, this finding enables us to apply the Ginzburg-Landau theory at high temperatures by changing the spacetime metric, potentially offering insights into achieving high-temperature superconductivity in these materials.

\end{abstract}

\maketitle

\section{INTRODUCTION}
Lorentz symmetry is a fundamental principle in physics, demanding that the laws governing the universe remain unchanged regardless of the laboratory's orientation or velocity through space  \cite{lor}.
 This symmetry plays a main role in both quantum field theory and Einstein's theory of relativity. Furthermore, it is closely linked to the principles of causation, as well as the symmetries of parity, charge, and time reversal \cite{k1,k2}.
The Lorentz group is a fundamental symmetry in particle physics while in condensed matter physics, the lattice breaks this symmetry of the vacuum, but in some lattices like graphene, the Lorentz symmetry appears in Dirac cones for fermions with Fermi velocity ($ v_F $) which is much less than the speed of light \cite{5}.
The Dirac materials have two cones with a crossing-point, where the energy of the valence and conduction bands are not equal anywhere, except at the zero dimensional Dirac points. As a result, the electrical conduction can be described by the movement of the fermions which is handled by the Dirac effective Hamiltonian,
\begin{equation}
H_D=v_F \gamma^0 \gamma^i p_i+mv_F^2\gamma^0,
\end{equation}
$ i=1...d $, where d is the space dimensions, and $ \gamma^\mu $ are Dirac matrices  where $ \gamma^0 $ represents time \cite{1}. Similar to the light cones in general relativity, the coaxial Dirac cones in solid state physics construct a flat spacetime at length scales much larger than the lattice constant with validity of Lorentz symmetry.
Although it is impossible to break the Lorentz symmetry of light cones, the Dirac cones 
can be tilted and tuned by a perpendicular electric field \cite{2}.
In tilted Dirac cones, the Lorentz symmetry is not valid like coaxial cones but the spacetime can be invariant under some deformation of the Lorentz group.
The organic compound $ \alpha-(BEDT-TTF)_2I_3 $ was the first discovered material  \cite{3} with tilted Dirac cones. 
The tilt of the Dirac cone in this compound is originated by interlayer magnetoresistance \cite{33}.

By tilting the Dirac cone, the Dirac effective Hamiltonian becomes,
\begin{equation}
H_D=v_F \gamma^0 \gamma^i p_i+mv_F^2\gamma^0+v_F\zeta^i p_i,
\end{equation}
where $\zeta^i$ is the tilt parameter.
This Hamiltonian is very interesting because it depends on the tilt parameter and as we see later, $\zeta^i$ defines the metric of the spacetime felt by the Dirac electrons. 
In other words, this is a curved spacetime world with a tunable curvature.
This property may enable us to find wonderful phenomena like high-temperature superconductivity at low pressures by tuning the tilt angle.
Since high-temperature superconductors are strongly restricted by an
extreme high-pressure condition ($\sim 100GPa$) that makes it
difficult to apply them in technology and industry.
The cuprate materials with $ T_c=133.5 \ K $ are considered as the
highest temperature superconductors at low pressures. 

 The Ginzburg-Landau (GL) model was the first superconductivity theory that could describe low temperature type-I superconductors without considering microscopic properties. This successful theory could predict some properties of 
  type-I superconductors as well.
After Bardeen-Cooper-Schrieffer (BCS) microscopic theory \cite{bcs}, a version of GL model was derived  by Gor'kov \cite{gor}.
In present work we intend to derive the general GL formalism for the tilted Dirac cone systems in curved spacetime, since the Minkowski metric is no longer valid.
For this purpose, we begin writing the BCS Hamiltonian according to the tilt parameter and corresponding metric by quantum field integrals, and after doing some algebra, finally,
  we obtain the GL action which depends on the tilt parameter. 
The next section is devoted to derivation of GL action in curved spacetime. It should 
be stated that we apply the approaches of Ref. \cite{alt} in our calculations.

\section{Method and Discussions }
The metric for Dirac matter is called Painlev\'e--Gullstrand
 \cite{pgu} with the form,
\begin{equation}
ds^2=-v^2_F dt^2+(d\bold{r}- v_F \bold{\zeta}dt)^2,
\end{equation}
where $ \bold{\zeta} $ encodes the tilt of the Dirac cone and $ v_F $ is the Fermi velocity which is considered $ 1 $ in our calculations and plays a role similar to the light speed $ (c) $. The Fermi velocity is usually two or three orders of magnitude smaller than c \cite{5}.
The explicit form of the above metric in 3+1 dimensions is,
\begin{equation}\label{met}
g_{uv}=
\begin{pmatrix}
\zeta^2-1 & -\zeta_x & -\zeta_y & -\zeta_z \\
-\zeta_x & 1 & 0 & 0\\
-\zeta_y & 0 & 1 & 0\\
-\zeta_z & 0 & 0 & 1
\end{pmatrix},
\end{equation}
where  $ \zeta^2= \zeta_x^2+\zeta_y^2+\zeta_z^2$ and the determinant of this metric is $ -1 $. 
The BCS Hamiltonian in style of quantum field theory for Dirac electron is,
\begin{eqnarray}\label{1}
H&=&\sum_{s}\int \sqrt{-g} \; d^dx \; \overline{\psi} _{s}\left[ i \gamma^a e^{\mu} _a (\partial _{\mu} +\Omega_{\mu}+g_{{\mu}b}A^b)-m\right.\nonumber\\
&-&\left.\Lambda+\phi^0 \right]\psi_{s}
-c_1\int \sqrt{-g} \;  d^dx \; \overline{\psi} _{\uparrow} \overline{\psi} _{\downarrow} 
\psi_{\downarrow} \psi_{\uparrow} .
\end{eqnarray}
where $ s $ is the electron spin, $g $ is the determinant of the metric tensor,
 $ \psi(x,\tau) $ is the Grassmann field, $e^{\mu} _a$ are the frame fields (in 3+1 are called vierbeins), $\Omega_{\mu}$ is the spin connection, $ A^b $ is the vector potential with $ A^0=0 $, $\Lambda $ is the Fermi kinetic energy, $m$ is the electron mass, $ \phi^0 $ is the scalar potential, and $ c_1 $ is the constant of the BCS Hamiltonian.
In our calculations, the constants $ e $, $ \hbar $, and $ c $ are considered $ 1 $.
The Grassmann field in Eq.\ref{1} is defined as,
\begin{equation}
\psi(x)=\sum_{s=1}^2 \int\frac{d^3p}{(2\pi)^3} \frac{1}{\sqrt{2\varepsilon_{\vec{p}} } }\left[ b^s_{\vec{p}} \; u^s(\vec{p}) \; e^{i\vec{p}.\vec{x}}   +  \overline{ c}^{s }_{\vec{p}} \; v^s(\vec{p}) \; e^{-i\vec{p}.\vec{x}}\right],
\end{equation} 
\begin{equation}
\overline{\psi}(x) =\sum_{s=1}^2 \int\frac{d^3p}{(2\pi)^3} \frac{1}{\sqrt{2\varepsilon_{\vec{p}} } }\left[\overline{b}^s_{\vec{p}} \; \overline{u}^s(\vec{p}) \; e^{-i\vec{p}.\vec{x}}   +   c^{s}_{\vec{p}} \; \overline{v}^s(\vec{p}) \; e^{i\vec{p}.\vec{x}}\right],
\end{equation} 
where the operators $ \overline{b}^s_{\vec{p}} $ and $\overline{ c}^{s }_{\vec{p}}$ create particles associated to the spinors $u^s(\vec{p})$ and $ \overline{v}^s(\vec{p})$ respectively, $ x $ and $ p$ refer to the momentum and the position of the Dirac electrons respectively and   $ \varepsilon_{\vec{p}}=\sqrt{m^2+p^2} $.

To use the further benefits of the field integrals, we write the quantum
 partition function in form of the coherent state path integral,
\begin{equation}
z=\int  D(\overline{\psi},\psi) \; e^{-S[\overline{\psi},\psi]},
\end{equation}
with the action,
\begin{eqnarray}\label{2}
S[\overline{\psi},\psi]&=& \sum_{s} \int  d^dx \left[\overline{\psi}_s ( \partial _{\tau} +
i \gamma^a e^{\mu} _a (\partial _{\mu} +\Omega_{\mu}+g_{{\mu}b}A^b)\right.\nonumber\\
&-&\left.m-\Lambda+i\phi^0 )\psi_{s} (x) -c_1 \overline{\psi} _{\uparrow} \overline{\psi} _{\downarrow} 
\psi_{\downarrow} \psi_{\uparrow} \right].
\end{eqnarray}
The conversion of $ \phi^0 $ into $ \partial _{\tau} + i \phi^0 $ originates the Wick rotation due to the imaginary time.

The term of the fields interactions ($ \overline{\psi} _{\uparrow} \overline{\psi} _{\downarrow} 
\psi_{\downarrow} \psi_{\uparrow} $)  in Eq.\ref{2} prevents the partition function from being calculated mathematically. 
Furthermore, in superconductivity phase, the electrons in adjacency of the Fermi surface are paired and behave as a boson, therefore we introduce a bosonic
field $\Delta$ to decouple the interaction. In fact $\Delta$ plays the role of the order parameter as well. The decoupling is done by a Hubbard-Stratonovich transformation as follows,
\begin{eqnarray}
 &exp&\left[c_1 \int d^dx  \overline{\psi} _{\uparrow} \overline{\psi} _{\downarrow} 
\psi_{\downarrow} \psi_{\uparrow}\right]=
\int  D(\overline{\Delta},\Delta)exp\left[\right.\nonumber\\
&-&\left.\int d^dx\left[\frac{1}{c_1}\vert \Delta \vert ^2
-(\overline{\Delta}\psi_{\downarrow} \psi_{\uparrow}+\Delta\overline{\psi} _{\uparrow} \overline{\psi}_{\downarrow})\right] \right],
\end{eqnarray}
with,
\begin{equation}
\Delta=\frac{g}{L^d}\sum_{k} {\left\langle {\Omega_{BCS}} \right|{\psi_{-k\downarrow}\psi_{k\uparrow}}\left| {\Omega_{BCS}} \right\rangle },
\end{equation}
\begin{equation}
\overline{\Delta}=\frac{g}{L^d}\sum_{k} {\left\langle {\Omega_{BCS}} \right|{{\overline{\psi}}_{k\uparrow}\overline{\psi}_{-k\downarrow}}\left| {\Omega_{BCS}} \right\rangle },
\end{equation}
where $L^d$ is the volume in $ d $ dimensions and
$ \left| \Omega_{BCS} \right\rangle  $ is the BCS wave function.

For convenient we use the Nambu spinor, 
\begin{equation}
\Psi=
\begin{pmatrix}
\psi _{\uparrow}\\
\overline{\psi} _{\downarrow}
\end{pmatrix}.
\end{equation}
Then the partition function reads,
\begin{eqnarray}
z=\int  D(\overline{\psi},\psi) \; \int D(\overline{\Delta},\Delta) \exp[ &-&\int d^dx  [\frac{1}{c_1}  \vert   \Delta \vert ^2\nonumber\\
&- &\overline{\Psi}(\Upsilon^{-1})\Psi]],
\end{eqnarray}

where,
\begin{align*}
\Upsilon=
\begin{pmatrix}
[G_{0p}]^{-1} & \Delta \\
\overline{\Delta} & [G_{0h}]^{-1}
\end{pmatrix},
\end{align*}
\begin{align*}
[G_{0p}]^{-1}= -\partial _{\tau} +
i \gamma^a e^{\mu} _a (\partial _{\mu} +\Omega_{\mu}+g_{{\mu}b}A^b)-m-\Lambda+i\phi^0,
\end{align*}
\begin{align*}
 [G_{0h}]^{-1}=-\partial _{\tau} 
-i \gamma^a e^{\mu} _a (\partial _{\mu} -\Omega_{\mu}+g_{{\mu}b}A^b)+m+\Lambda-i\phi^0.
\end{align*}
$ \Upsilon $ is the Gor'kov Green function, $ p $ and $ h $ refer to the particle and hole,
$[G_{0p}]^{-1}$ and $  [G_{0h}]^{-1} $ represent the non-interacting Green functions of the particle and hole.

By using the following identity for a typical Grassmann field $ \phi $ and operator $A$,
\begin{equation}
\int  D(\overline{ \phi},\phi) \; e^{-(\overline{ \phi} A \phi)}=det \; A,
\end{equation}
we can write the partition function as,
\begin{equation}
z= \int D(\overline{\Delta},\Delta)exp\left[-\int d^dx\frac{1}{c_1}\vert \Delta \vert ^2 + ln \; (det \; \Upsilon^{-1})\right].
\end{equation}

By applying the identity,
\begin{equation}
 ln \; (det \; A)= tr \; (ln \; A ),
\end{equation}
the partition function reads,
\begin{equation}
z= \int D(\overline{\Delta},\Delta)exp\left[-\int d^dx\left(\frac{1}{c_1}\vert \Delta \vert ^2\right) + tr \; \left(ln \; \Upsilon^{-1}\right)\right].
\end{equation}

%

\begin{figure}
\includegraphics[scale=0.7]{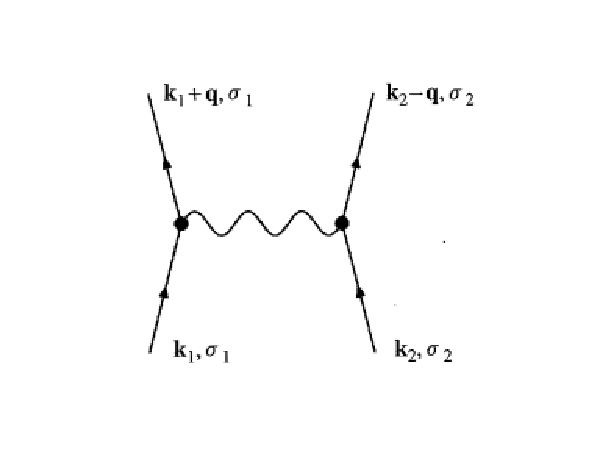}
\caption{Interaction of electrons via exchange
of a boson at the Fermi surface
due to the exchange of crystal lattice phonons.
$ k $ and $\sigma$ are wave vector and spin of the electron, respectively.}\label{inter}
\end{figure}

Near the phase transition, the gap parameter $ \Delta $ is 
small in comparison with the temperature, therefore we are able to use perturbation theory. The alternative interesting case that makes it possible to use the perturbation expansion in higher temperatures, is tilting the Dirac cone to change the metric in a way that $ \Upsilon^{-1}\simeq \Upsilon^{-1}_0$. This case refers to the high-temperature superconductivity occurring by tilting the Dirac cone.
Therefore by defining $ \Upsilon^{-1}_0\equiv   \Upsilon^{-1}\vert_{\Delta=0}$ and 
 $ \stackrel{\wedge}{\Delta }\equiv \begin{pmatrix}
0 & \Delta \\
\overline{\Delta} & 0
\end{pmatrix} $
we have,
\begin{eqnarray}\label{213}
 tr \; (ln \; \Upsilon^{-1})&=& tr \; \left[ln \; \left[ \Upsilon^{-1}_0(1+\Upsilon_0 \stackrel{\wedge}{\Delta })\right]\right]= \underbrace{ tr \; [ ln \; \Upsilon^{-1}_0}_{const.}]\nonumber\\
 &+& tr \; \left[ln \; (1+\Upsilon_0 \stackrel{\wedge}{\Delta })\right].
\end{eqnarray}
The constant term is inessential for phase transition since it refers to the free energy of non-interacting electron gas. Expanding the last term of Eq.\ref{213}, we get,
\begin{equation}\label{589}
 tr \; \left[ln \; \left(1+\Upsilon_0\stackrel{\wedge}{\Delta }\right)\right]=-\sum^{\infty }_{n=0} \frac{1}{2n} \; tr\left(\Upsilon_0\stackrel{\wedge}{\Delta }\right)^{2n}.
\end{equation}
Since the Hamiltonian is real, the odd terms vanish.
For $ n=1 $,
\begin{eqnarray}
 &tr& \left[ln \; \left(1+\Upsilon_0 \stackrel{\wedge}{\Delta }\right)\right]=- \frac{1}{2} \; tr\left(\Upsilon_0\stackrel{\wedge}{\Delta }\right)^{2}=\nonumber\\
 &-& \frac{1}{2} \; tr
 \begin{pmatrix}
 G_{0p}G_{0h}\Delta\overline{\Delta} & 0 \\
0 & G_{0p}G_{0h}\Delta\overline{\Delta} 
\end{pmatrix} \nonumber\\
&=&-G_{0p}G_{0h}\Delta\overline{\Delta} .
\end{eqnarray}
Transforming to the momentum space,
\begin{equation}\label{591}
 tr \; \left[ln \; \left(1+\Upsilon_0 \stackrel{\wedge}{\Delta }\right)\right]=-\sum_{q} \frac{T}{L^d} \sum_{p} G_{0,p}G_{0, -p+q}\Delta\overline{\Delta},
\end{equation}
where $ q $ is the phonon momentum which interacts between the two electrons of cooper pair, see Fig.\ref{inter}.

The second order term of the action reads,
\begin{equation}\label{s2}
S^{(2)}[\Delta,\overline{\Delta}]=\sum_{q}\Gamma_q^{-1} \vert \Delta\vert^2,
\end{equation}
\begin{equation}
\Gamma_q^{-1}=\frac{1}{c_1}- \frac{T}{L^d} \sum_{p} G_{0,p}G_{0, -p+q},
\end{equation}
%
%
where
$ \Gamma_q^{-1} $ is the vertex function, and
$ \vert \Delta\vert^2=\overline{\psi} _{\uparrow} \overline{\psi} _{\downarrow} 
\psi_{\downarrow} \psi_{\uparrow} $ is the four point correlation function which demonstrates the probability of constitution of cooper pair. 
We are now in a position to explore the consequences of the symmetry 
breaking. 
$ \Gamma_q^{-1} $ is the only term in Eq.\ref{s2} which corresponds to a sign change of the quadratic action.
 Therefore, near critical point as $\Gamma_q^{-1} \rightarrow 0$, it must behave like,
\begin{equation}\label{eq102}
 \Gamma_q^{-1}  \sim T-T_c.
\end{equation}

Now we intend to justify the scale relation (\ref{eq102}). 
It is straightforward to write the pair correlation function $ Q_p $ as a function of 
the Fermi-Dirac occupation number $ n_F(\epsilon) $,
 \begin{eqnarray} \label{tld}
 Q_q&\equiv & \frac{T}{L^d} \sum_{p} G_{0,p}G_{0, -p+q}\nonumber\\
 &=&\frac{1}{L^d} \sum_{p}\frac{1-n_F(\epsilon_{p})
 -n_F(\epsilon_{-p+q})}{
- \epsilon_{p}-\epsilon_{-p+q} },
\end{eqnarray}
For more information see Ref.\cite{alt}
By applying the identity,
\begin{equation}
\frac{1}{L^d}\sum_{p}H(\epsilon_p)=\int d\epsilon \nu (\epsilon)H(\epsilon),
\end{equation}
to Eq.\ref{tld} for constant $ \Delta(q=0) $, it becomes, 
\begin{equation}\label{tld2}
- \frac{T}{L^d} \sum_{p} G_{0,p}G_{0, -p}=\int d\epsilon \;  
\; \nu(\epsilon) \frac{1-2n_F(\epsilon,T)}{2\epsilon},
\end{equation}
 where $ H(\epsilon) $ is a typical function of single particle energy state $ (\epsilon) $, $\nu(\epsilon)$ is the density of states, and $ \epsilon_p=\epsilon_{-p} $ . The integration band in Eq.\ref{tld2} is limited to $ \epsilon \simeq \epsilon_F $, therefore 
 $\nu(\epsilon)\simeq \nu$.
By using the Tylor expansion of $n_F(\epsilon,T)$ near the critical point,
\begin{eqnarray}
n_F(\epsilon,T)&\simeq & n_F(\epsilon,T_c)+(T-T_c)\partial_T\vert_{T=T_c}n_F(T)\nonumber\\
&=&-\partial_\epsilon n_F(\epsilon,T_c)(T-T_c)\frac{\epsilon }{T},
\end{eqnarray}
thus,
\begin{eqnarray}\label{109}
- \frac{T}{L^d} \sum_{p} G_{0,p}G_{0, -p+q}&\simeq &\frac{1-2\left[-\partial_\epsilon n_F(\epsilon,T_c)(T-T_c)\frac{\epsilon }{T}\right]}{2\epsilon}\nonumber\\
&=&\frac{1}{2\epsilon}+\frac{\partial_\epsilon n_F(\epsilon,T_c)(T-T_c)}{T}\nonumber\\
&=&\frac{1}{2\epsilon}+\frac{\sim\beta^{-1} (T-T_c)}{T}\nonumber\\
&\sim & T-T_c.
\end{eqnarray}

In ordinary GL formalism, at $T \lesssim T_c$ the $ S^{(2)} $ action only can not justify the phase transition, therefore the existance of
 $ S^{(4)} $ term is necessary in calculations. 
According to Eqs.\ref{589} and \ref{591}, the $ S^{(2n)} $ action is,
\begin{eqnarray}
S^{(2n)}&=&\frac{1}{2n} tr\left( \Upsilon_0\stackrel{\wedge}{\Delta }\right)^{2n}\nonumber\\
&=&\frac{(-1)^n}{2n} \frac{T}{L^d} \sum_{p,q} \left(G_{0,p}G_{0, -p+q}\right)^n\vert\Delta\vert^{2n}
\end{eqnarray}
Now we intend to enter the spatial gradient $(\partial \Delta )$ to GL action.
For simplicity we use the flat metric at first and
 expand the pair correlation function (Eq. \ref{tld} ) as follows,
\begin{eqnarray}\label{8368}
Q_p&=&- \int \frac{d^dp}{(2\pi)^d}\frac{1-n_F(\epsilon_{p+\frac{q}{2}})
 -n_F(\epsilon_{p-\frac{q}{2}})}{
- \epsilon_{p+\frac{q}{2}}-\epsilon_{p-\frac{q}{2}} }\nonumber\\
&\simeq & - \int \frac{d^dp}{(2\pi)^d}\frac{1-2n_F(\epsilon_p)-\partial^2_\epsilon n_F(\epsilon_p)\frac{(p.q)^2}{4m^2} }{2\epsilon_p}\nonumber\\ 
&=& Q_0-\frac{\mu q^2}{12m} \int d \epsilon \; \epsilon^{-1}\partial^2_\epsilon n_F(\epsilon_p)\nonumber\\ 
&=& Q_0-\frac{c_2v_F^2q^2}{24T^2},
\end{eqnarray}
where $\mu $ is the chemical  potential, $ \epsilon_{p+q} \simeq \epsilon_p+\frac{p.q}{m} $, $c_2$ is a numerical constant, and $ \frac{p^2}{2m}\simeq \mu $.
$ q^2 $ in Eq.\ref{8368} is the Fourier transform of $\partial^2$, 
therefore by inserting the correlation function into  Eq.\ref{s2}, the spatial gradient $(\partial \Delta )$ appears in the second order term of the action,
\begin{equation}
S^{(2n)}\sim  \int d^dx \left[ \vert\Delta \vert^{2n} + \vert \partial \Delta \vert ^2 + \cdots\right] .
\end{equation}
At orders $n>2 $ of the expansion, spatial gradient can be neglected 
due to $\Delta\ll T$.  

For finding the relativistic GL formalism, we should write the normal derivatives in a covariant form of general relativity.
For this purpose, the Christoffel symbols are defined by Eq.\ref{met},
\begin{equation}\label{700}
\Gamma^\mu_{\nu \rho}=\frac{1}{2}g^{\mu \eta }(\partial_\nu g_{\rho \eta}+\partial_\rho g_{\nu \eta}-\partial_\eta g_{\nu \rho} ).
\end{equation}
and the covariant derivatives for a typical vector $F_{\nu}$ is given by,
\begin{equation}\label{701}
\bigtriangledown_\rho F^{\nu}=F^\nu_{,\rho}+\Gamma^\nu_{\mu \rho} F^{\mu}.
\end{equation}
By replacing the normal derivatives of pair correlation function with covariant ones, the relativistic GL formalism for anisotropic order parameter $ (\Delta^\nu) $ becomes, 
\begin{equation}\label{fin}
S^{(2n)} \sim \int  d^dx \left[ \; \vert\Delta^\nu \vert^{2n} +  (\bigtriangledown_\rho+\Omega_\rho) \Delta^\nu + \cdots\right].
\end{equation}
where the non-zero Christoffel symbols are $ \Gamma^0_{00}, \Gamma^1_{01}, \Gamma^2_{22},$ and $\Gamma^3_{03} $.

 The BCS theory explains the low-temperature superconductivity based on the condition $\Delta^\nu\ll T$, but in present work (tilted Dirac cone materials) this condition is still applicable at high-temperature as well due to the tilt of the Dirac cone. The metric dependency of relation \ref{fin} makes it possible to respect the BCS condition ($\Delta^\nu\ll T$) even for $\Delta^{\nu}< T$ or $\Delta^{\nu} \sim T$ cases by changing the metric of spacetime. 
 By approaching the tilt parameter to zero, the spin connection approaches zero and $ g_{uv} $ becomes a Minkowski metric. 
 According to this interesting result, it would be probable to find high-temperature superconductivity at low pressures in tilted Dirac cone systems just by changing the external electric field.
 

\section{Summary and Concluding remarks}

The problem of finding high-temperature superconductors has been one of the main challenges in condensed
matter physics. Many attempts have been carried out
to observe high-temperature superconductivity, however, in
most of them the system requires an extreme high-pressure condition.
The other problem is the absence of a general theory to explain the high-temperature superconductivity.
 In present work, we have derived a general form of Ginzburg-Landau theory based on the BCS Hamiltonian in curved spacetime and has applied it for the tilted Dirac cone materials with a non-flat metric. 
 Curved spacetime is a basic concept in cosmology and astrophysics with extreme conditions like too heavy masses or high energies. Meanwhile, Dirac materials in condensed matter physics make it possible to observe the general relativity of Einstein easily just by applying an electric field on these materials to tilt the Dirac cone and change the metric. 
 This property enables us to apply the GL formalism at high temperatures and may help finding high-temperature superconductivity in such systems in future.
The tilt parameter of Dirac cones is determined by a perpendicular external electric field; therefore, it is very interesting that we are able to tune the metric of a system by changing the electric field, in velocities much smaller than the speed of light.
In BCS theory, the main restriction is the order parameter to be much smaller than the  temperature of the system ($\Delta^{\nu}\ll T$) while in systems with tilted Dirac cone, the order parameter depends on temperature and the metric of the spacetime.
This dependency makes it possible to respect the  BCS condition ($\Delta^\nu\ll T$) by changing the metric, and to find higher critical temperatures of superconductivity by changing the direction or magnitude of the electric field in tilted Dirac cone materials. 

\section{Declarations}
This research did not receive any grant from funding agencies.
\subsection{Conflict of interest}
 The author declares that he has no conflict of
interest or competing interests.
\subsection{Data availability statement}
 All data generated or analysed during this study are included in this published article.



\end{document}